\documentclass[mathleft
]{an}
\usepackage{graphicx}
\usepackage{times}
\begin{document}

\Pagespan{789}{}
\Yearpublication{2010}%
\Yearsubmission{2009}%
\Month{12}%
\Volume{999}%
\Issue{88}%

\title{Masses and Luminosities of O and B - type stars and red super giants}

\author{M.M. Hohle\inst{1,2}\fnmsep\thanks{Corresponding author:
  {mhohle@astro.uni-jena.de}}
\and  R. Neuh\"auser\inst{1}
\and  B.F. Schutz\inst{3,4}
}
\titlerunning{Masses and Luminosities of O and B - type stars and red super giants}
\authorrunning{Hohle et al.}
\institute{
Astrophysical Institute and University-Observatory Jena, Schillerg\"asschen 2-3, 07745 Jena, Germany 
\and 
Max-Planck-Institute for Extraterrestrial Physics, Giessenbachstrasse, 85741 Garching, Germany
\and
Max-Planck-Institute for Gravitational Physics Potsdam, Am M\"uhlenberg 1, 14476 Golm, Germany
\and
School of Physics and Astronomy, Cardiff University, 5, The Parade, Cardiff, UK, CF24 3AA
}
\received{}
\accepted{}
\publonline{later}


\abstract{
Massive stars are of interest as progenitors of super novae, i.e. neutron stars and black holes, which can be sources of gravitational waves. Recent population synthesis models can predict neutron star and gravitational wave observations but deal with a fixed super nova rate or an assumed initial mass function for the population of massive stars.\\
Here we investigate those massive stars, which are supernova progenitors, i.e. with O and early B type stars, and also all super giants within 3kpc. We restrict our sample to those massive stars detected both in 2MASS and observed by Hipparcos, i.e. only those stars with parallax and precise photometry.\\
To determine the luminosities we calculated the extinctions from published multi-colour photometry, spectral types, luminosity class, all corrected for multiplicity and recently revised Hipparcos distances. We use luminosities and temperatures to estimate the masses and ages of these stars using different models from different authors.\\
Having estimated the luminosities of all our stars
within 3kpc, in particular for all O- and early B-type stars,
we have determined the median and mean luminosities for
all spectral types for luminosity classes I, III, and V.  
Our luminosity values for super giants deviate from earlier results: Previous work generally overestimates distances and luminosities compared to our data, this is likely due to Hipparcos parallaxes (generally more accurate and larger than previous ground-based data) and the fact that many massive stars have recently been resolved into multiples of lower masses and luminosities.\\
From luminosities and effective temperatures we derived masses and ages using mass tracks and isochrones from different authors. From masses and ages we estimated lifetimes and derived a lower limit for the supernova rate of $\approx20$ events/Myr averaged over the next 10~Myrs within 600~pc from the sun. These data are then used to search for areas in the sky with higher likelihood for a supernova or gravitational wave event (like OB associations).
}
\maketitle

\section{Introduction}

To estimate the ages and masses of stars,
one almost always needs their luminosities and
temperatures to compare their location in the
H-R diagram with theoretical isochrones
and tracks. Only in a few rare cases, other
mass (or age) estimates are possible, e.g.
in eclipsing double-lined binaries.
Luminosity, mass, and age are very important
parameters to study and understand the
formation of stars. In particular for massive
stars, as studied here, the formation mechanism
is still a matter of debate, either accretion
from massive disks and/or coagulation of lower-
mass stars (see e.g. Zinnecker \& Yorke, 2007,
for a recent review).\\
For a lot of studies, typical mean luminosities and masses
of stars of a particular spectral type and
luminosity class (LC) are necessary, e.g. spectro-photometric distance or mass - luminosity relation.\\
Here, we use Hipparcos (Perryman et al., 1997)
parallaxes to re-estimate the luminosities of all massive
stars within 3kpc, for which both new Hipparcos (van Leeuwen, 2007a, b) and 2MASS (Cutri, 2003)
data are available. We use Hipparcos/Simbad (BV) and
2MASS (JHK) photometry together with the known
spectral type and luminosity class to estimate the
extinction. From these data, we also estimate
all luminosities and masses.\\
We restrict our sample to those stars which are
assumed to be progenitors to supernova and/or
neutron stars. Data as determined in our study are also
necessary ingredients to population synthesis models
to explain current neutron stars observations
and to predict future gravitational wave detections.

\section{The sample}

\begin{table*}
\centering
\caption{Input data of the first ten stars sorted by ascending relative error of the parallaxes (new reduced Hipparcos parallaxes from van Leeuwen, 2007, corrected using equation 21 in Smith \& Eichhorn, 1996). B and V band magnitudes and spectral types are obtained from the Simbad data base (Hipparcos), JHK magnitudes and their errors derived from the 2MASS catalogue. If the spectral type is listed in  Pourbaix et al. (2007) we give this value. For conversion from spectral type and luminosity class to temperature see section 4. The complete sample will be available at the ADS data base in electronic form.}
\label{initVal}
\begin{tabular}{cc|ccccc|c|c|c}
\hline
&Hip& B &	V & J &	H & K &	$\pi$ & SpType& $\mathrm{T_{eff}}$\\
  &   & \multicolumn{5}{c|}{[mag]}&	[mas] & & [K]\\

\hline

1 & 30122 & 2.83 & 3.00 & 3.464$\pm$0.260 & 3.503$\pm$0.242 & 3.670$\pm$0.272 & 9.09$\pm$0.13 & B3V & 18700 \\
2 & 86414 & 3.636 & 3.794 & 4.267$\pm$0.200 & 4.349$\pm$0.258 & 4.228$\pm$0.016 & 7.24$\pm$0.13 & B3IV & 17900 \\
3 & 39138 & 4.648 & 4.797 & 5.161$\pm$0.037 & 5.259$\pm$0.034 & 5.260$\pm$0.018 & 6.60$\pm$0.13 & B3V & 18700 \\
4 & 97278 & 4.278 & 2.724 & 0.276$\pm$0.168 & -0.544$\pm$0.224 & -0.720$\pm$0.244 & 8.34$\pm$0.17 & K3II & 4140 \\
5 & 69996 & 3.375 & 3.536 & 3.970$\pm$0.228 & 3.893$\pm$0.210 & 4.102$\pm$0.288 & 9.75$\pm$0.20 & B2.5IV & 19525 \\
6 & 99473 & 3.197	& 3.242 & 3.293$\pm$0.232 & 3.278$\pm$0.196 & 3.295$\pm$0.214 & 11.50$\pm$0.24 & B9II & 11000 \\
7 & 76600 & 3.490	& 3.644 & 3.990$\pm$0.230 & 4.046$\pm$0.212 & 4.120$\pm$0.027 & 8.98$\pm$0.20 & B2.5V & 20350 \\
8 & 67464 & 3.190 &	3.390 & 4.014$\pm$0.260 & 4.139$\pm$0.222 & 4.240$\pm$0.288 & 7.55$\pm$0.17 & B2V & 22000 \\
9 & 79404 & 4.421 &	4.567 & 4.980$\pm$0.250 & 5.010$\pm$0.029	& 4.976$\pm$0.027 & 6.88$\pm$0.16& B2V & 22000 \\
10 & 32759 & 3.402 & 3.515 &	3.804$\pm$0.274 &	3.679$\pm$0.252 &	3.547$\pm$0.258 & 5.00$\pm$0.12 & B1.5IV & 22675 \\

\hline
\end{tabular}
\end{table*}

\begin{figure}
  \centering
   \resizebox{\hsize}{!}
{
   \includegraphics[viewport=100 260 500 550, width=0.48\textwidth]{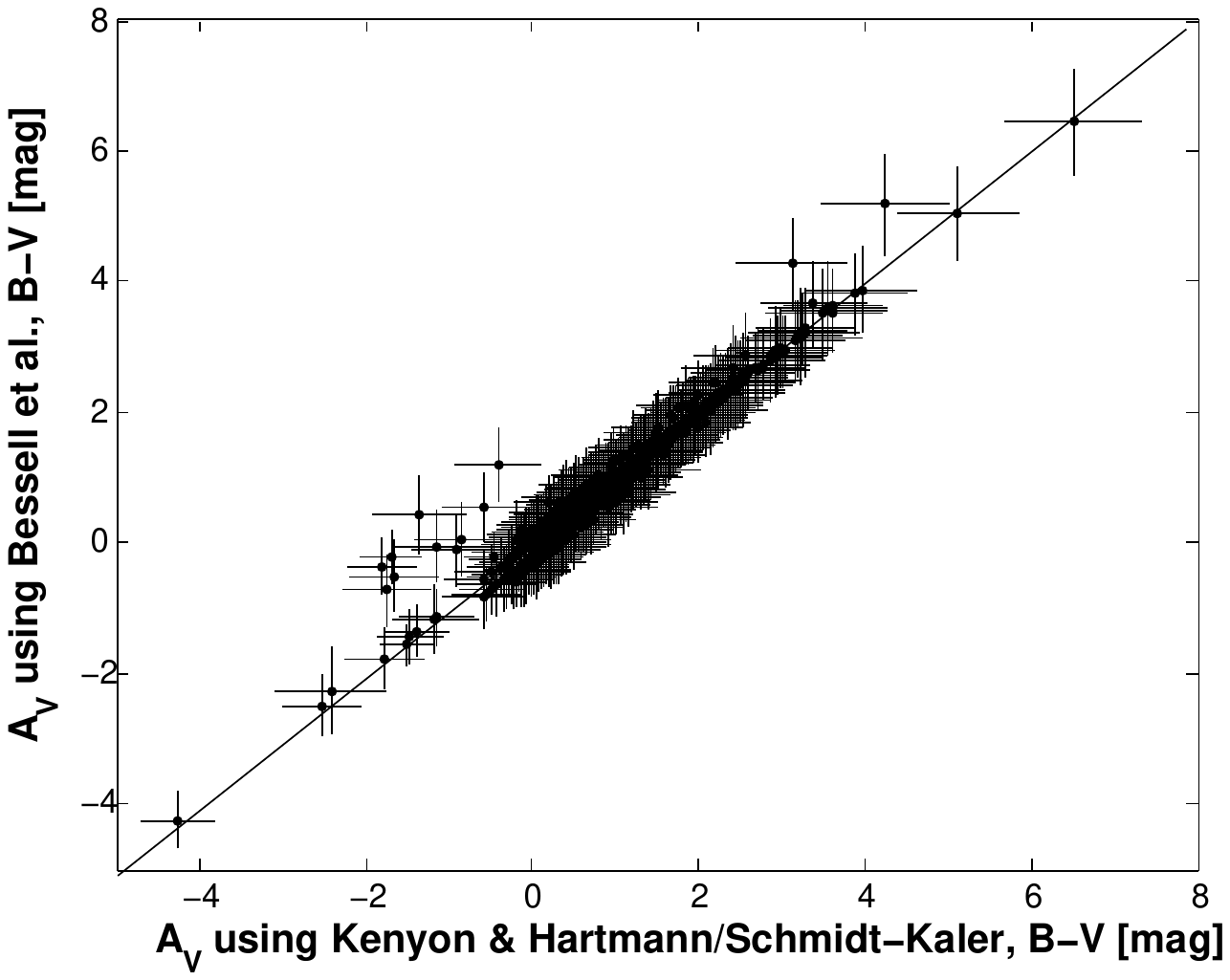}
}
   \caption{$A_{V}$ values from single stars calculated from $(B-V)$ using the intrinsic colours from different authors. Errors denote to $1\sigma$.}
              \label{B_V}
    \end{figure}

We compile a list of all known massive stars, which are supposed to explode as supernova (SN), i.e. for LC V and IV spectral types equal or earlier than B4, for LC III equal or earlier than B9 and for LC I and II all spectral types (massive red giants and super giants), all within a distance to the sun of 3kpc. This distance is chosen, so that we are complete for stars earlier than B3V with $A_{V}\leq$2.5~mag (limit for Hipparcos with at least 1~mas accuracy) and for comparison with the population synthesis in Popov~et~al. (2005). This list contains 16304 stars selected from Simbad, 3042 of them have revised Hipparcos parallaxes (van Leeuwen, 2007) and 2713 of those Hipparcos stars also have JHK magnitudes in 2MASS (Two {\bf M}icron {\bf A}ll {\bf S}ky {\bf S}urvey, Cutri, 2003), searching by the Hipparcos identifier. Some stars (for example Hip~22392 or Hip~23527) from the Magellanic Clouds accidently have parallaxes $\geq$0.33~mas listed in Hipparcos and/or in van Leeuwen (2007) and would be in this sample. Therefore we cut out circular regions with a radius of seven degrees for the Large Magellanic Cloud and 3.5 degrees for the Small Magellanic Cloud. Finally 2668 stars from the original 2713 stars (Hipparcos and 2MASS) are left in our list.\\
We have checked all stars for information about multiplicity in Simbad and catalogues about spectroscopic and eclipsing binaries, namely the binary catalogues from Pourbaix et al. (2007) and Docobo \& Andrade (2006) (both for spectroscopic binaries) and from Bondarenko \& Perevozkina (1996), Brancewicz \& Dworak (1980)\footnote{For Hip~108317 with $\sim20yrs$ orbit period, Brancewicz \& Dworak (1980) did not obtain dynamical masses, so that we obtain and use own masses for both components from public data.}, Surkova \& Svechnikov (2004) and Perevozkina (1999) listing eclipsing binaries. For 302 stars, there is not enough data available on the companion(s) to estimate parameters like luminosity correctly for all
components, so that we omit them from our list. Our list then contains 2323 (247 multiples + 2076 singles, after checking for redundancy) stellar systems, with multiples counted once, with a total of 2398 stars having all parameters for the mass calculation.\\
There are 247 spectroscopic or eclipsing systems in our list from the papers mentioned above. All those papers (expect Pourbaix et al., 2007) list dynamical masses, which are better than our model-dependent masses, so that we will use the published dynamical masses; for the stars in Pourbaix et al. (2007), good photometry is given for all known components, so that we can estimate the masses for all components from the published data (as we do for all other stars in our list).\\
The input data are listed in Table~\ref{initVal}. 

\begin{figure}[t]
  \centering
   \resizebox{\hsize}{!}
{
   \includegraphics[viewport=100 260 500 550, width=0.48\textwidth]{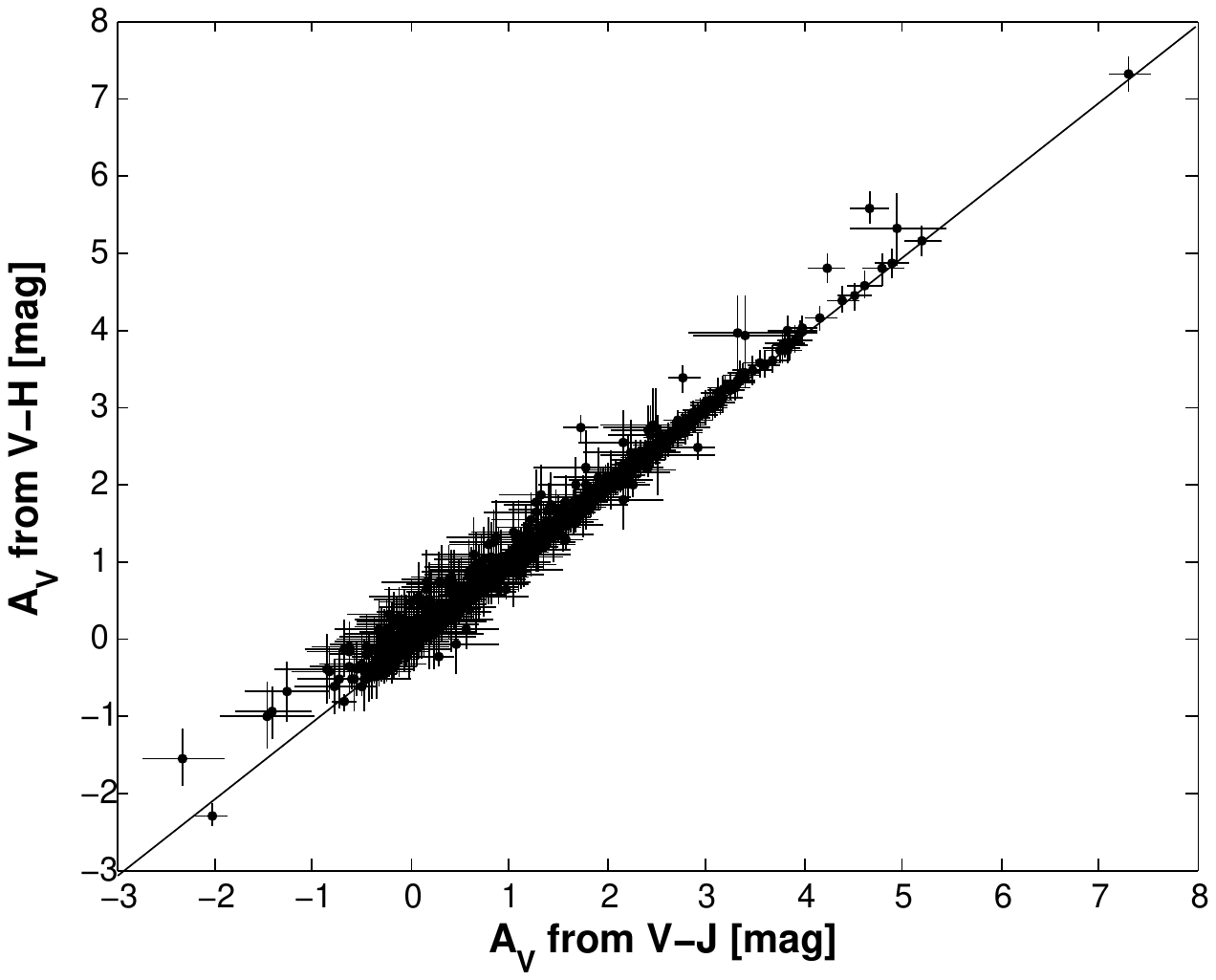}
}
   \caption{$A_{V}$ values from single stars calculated from $(V-J)$ and $(V-H)$ using the intrinsic colours listed in Kenyon \& Hartmann (1995)/Schmidt-Kaler (1982). Errors denote to $1\sigma$.}
              \label{KH}
    \end{figure}

\section{Colours and extinction}

\begin{figure}
  \centering
   \resizebox{\hsize}{!}
{
   \includegraphics[viewport=100 260 500 550, width=0.48\textwidth]{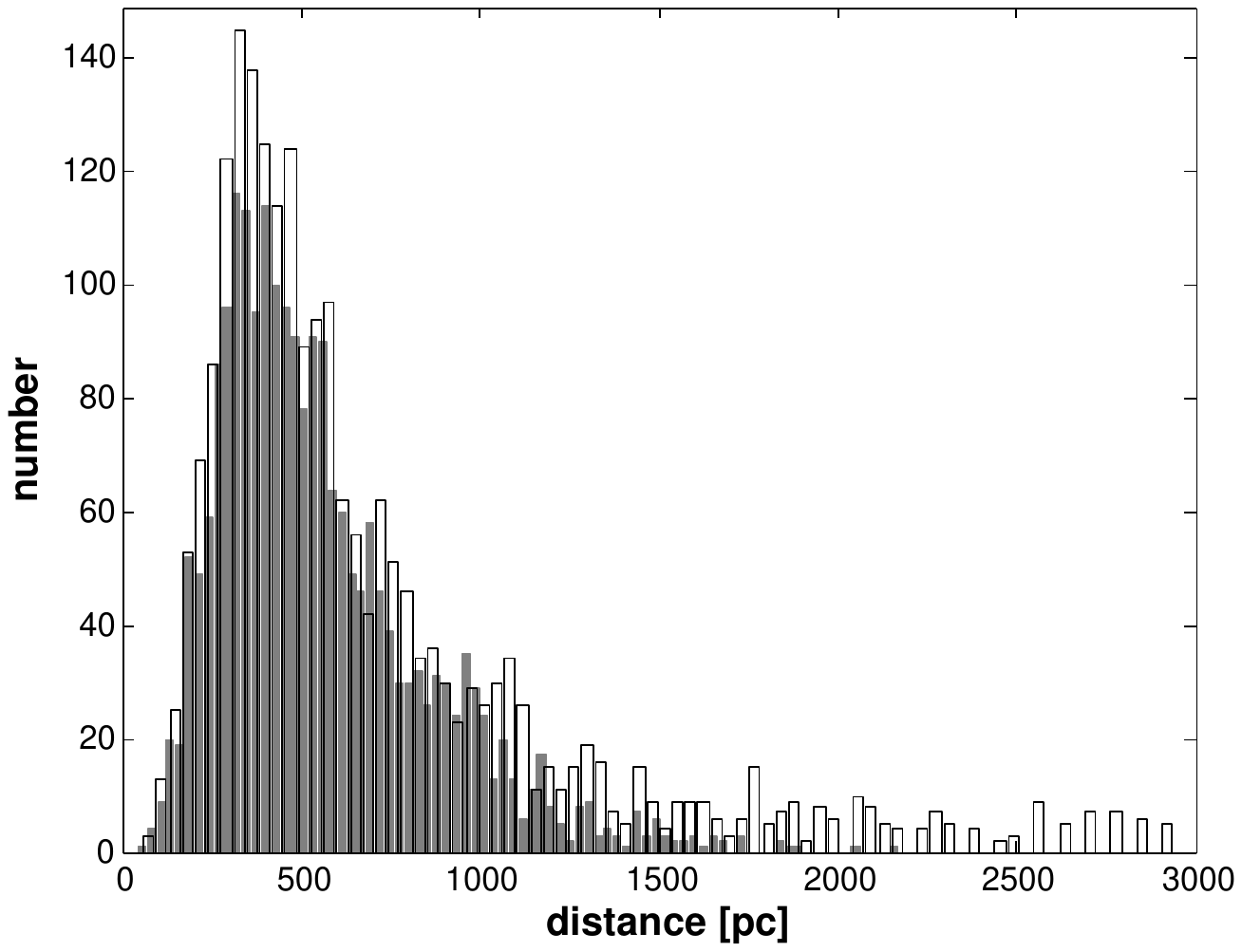}
}
   \caption{Histogram of the new Hipparcos distances (van Leeuwen, 2007) from all 2323 stars in the final sample (white bars) and after the application of the Smith \& Eichhorn (1996) correction (grey), see also Figure~\ref{plxcorr}.}
              \label{dist}
    \end{figure}

\begin{figure}
  \centering
   \resizebox{\hsize}{!}
{
   \includegraphics[viewport=100 260 500 550, width=0.48\textwidth]{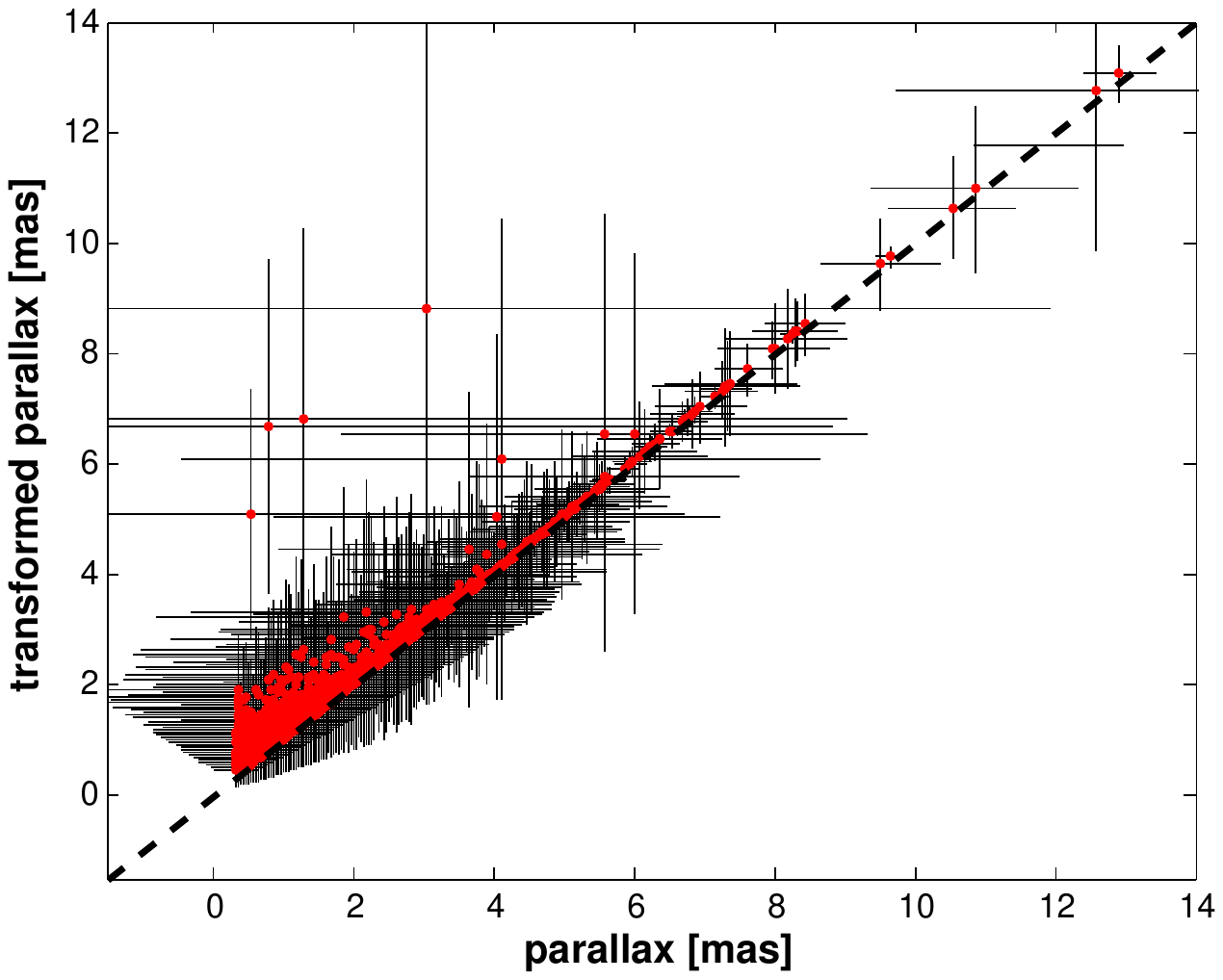}
}
   \caption{ Transformation of the Hipparcos parallaxes (van Leeuwen, 2007) applying the Smith \& Eichhorn (1996) correction (red dots) shown with 1$\sigma$ error bars from all 2323 stars in the final sample. The one-to-one relation is indicated as dashed line.}
              \label{plxcorr}
    \end{figure}

\begin{figure}
  \centering
   \resizebox{\hsize}{!}
{
   \includegraphics[viewport=100 260 500 550, width=0.48\textwidth]{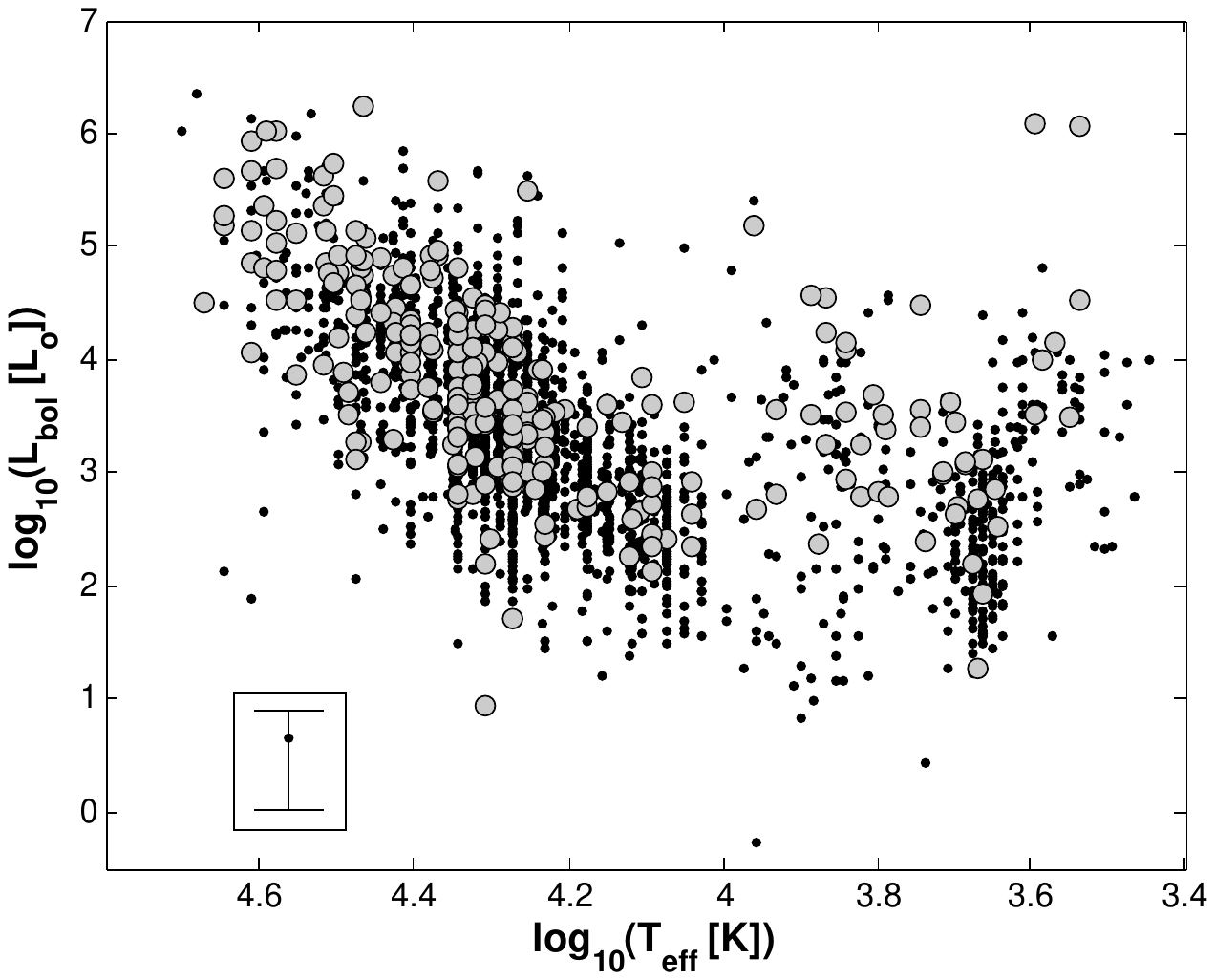}
}
   \caption{H$-$R diagram of the 2323 stars (dots) in the final list including 247 resolved massive multiples (grey circles). The large scatter around the main sequence results from the parallax errors. The luminosities were calculated after applying the Smith \& Eichhorn (1996) correction. A few binaries appear below the main sequence, but are consistent to being main sequence within the errors for the luminosities (a representative error bar is shown in the box).}
              \label{HRD}
    \end{figure}

The 2MASS magnitudes are well measured, with a median of the relative error of 0.34\% for J magnitudes. More than 96\% of all J magnitudes have errors lower than 10\%. The errors of the H and K magnitudes are comparable to those from J (Cutri, 2003). We use a general error of 1\% for the B and V magnitudes (Simbad and Hipparcos). The calculation of the bolometric corrections from the spectral types and the extinction ($A_{V}$) due to the interstellar medium follows the procedure in Hohle et al. (2009) using the bolometric corrections and the intrinsic colour indices from Bessell et al. (1998), Kenyon \& Hartmann (1995) and Schmidt-Kaler (1982). For the spectral types M4-6, we cannot use Bessell et al. (1998), which goes down to 3500K only, so that we use only Kenyon \& Hartmann (1995) and Schmidt-Kaler (1982) for these stars.\\
We calculated the $A_{V}$ values from BVJHK colours of the single stars from the final list of 2323 stars and fit them to the one-to-one relation ($Y(x)=Ax+a$) with the results listed in Table \ref{phot_single}. We only use those colours for our calculations, which are bold faced in Table \ref{phot_single}. The criteria for selection is the following: If one linear fit is not consistent to $\geq2$ others, we do not use it, for example $(B-V)_{0}$ from Kenyon \& Hartmann (1995). We treat a linear fit as consistent to another one, if $A=1\pm0.1$ considering the scattering $dA$ and if $a=0\pm0.1$mag considering $da$. With this criteria we select $(V-K)_{0}$, $(V-J)_{0}$ and $(V-H)_{0}$ from Kenyon \& Hartmann (1995)/Schmidt-Kaler (1982). One exception is $(V-K)_{0}$ from Bessell et al. (1998). Although it is not consistent to more than two linear fits from Bessell et al. (1998), it is consistent to $(V-K)_{0}$ from Kenyon \& Hartmann (1995), whose consistency is already shown.\\

\begin{figure*}
  \centering
   \resizebox{\hsize}{!}
{
   \includegraphics[viewport=-20 0 550 750, width=0.1\textwidth]{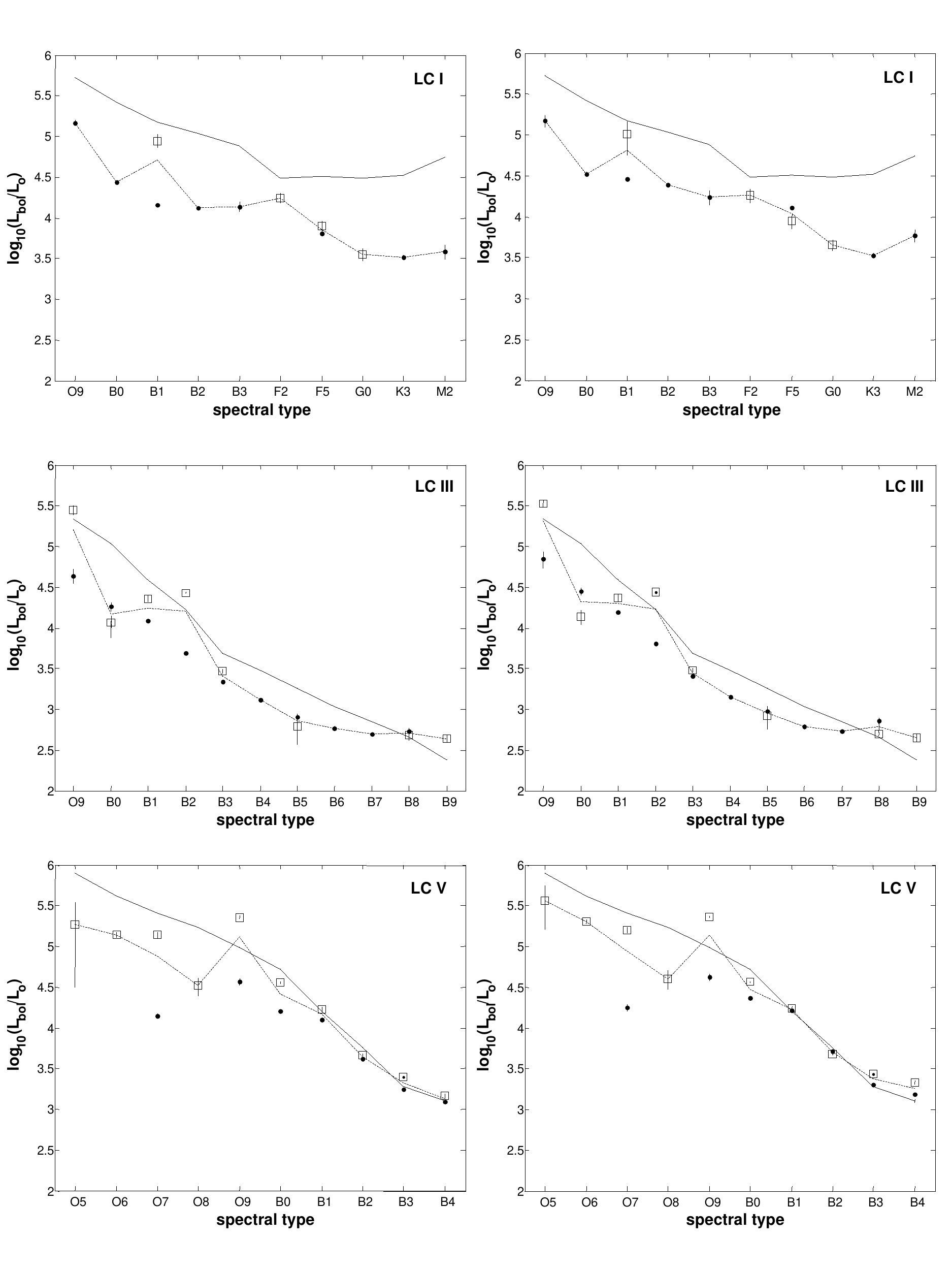}
}
   \caption{Median bolometric luminosities after the application of the Smith \& Eichhorn (1996) correction (left panel) for their parallax (dots, at least five stars per spectral sub-type) and the median luminosities of the stars from the catalogue of Pourbaix et al. (2007) (squares, at least three stars per spectral sub-type), the dotted lines show the linear interpolation between the two subsamples compared to the standard bolometric luminosities from Schmidt-Kaler (1982) as solid lines. The error bars give the standard deviations (in some cases the error bars are smaller than the symbol size). Right panel: same without Smith \& Eichhorn (1996) parallax correction. Schmidt-Kaler (1982) overestimates the luminosities due to ground based parallaxes and unresolved multiples.}
              \label{lc}
\end{figure*}

\begin{figure*}
  \centering
   \resizebox{\hsize}{!}
{
   \includegraphics[viewport=-20 0 550 750, width=0.1\textwidth]{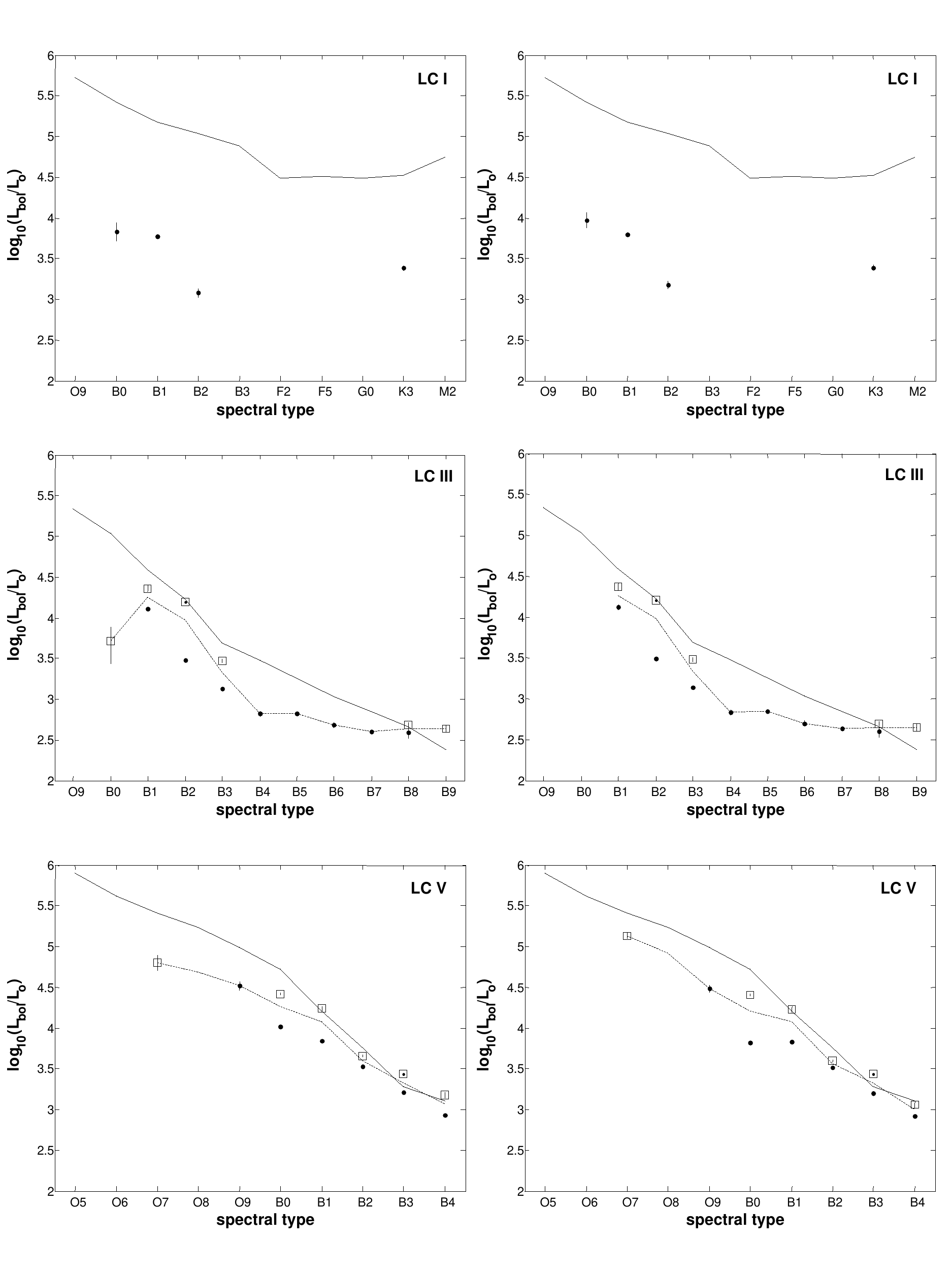}
}
   \caption{Same as in Figure~\ref{lc}, but only for those stars within 600~pc. Note that in this case the number of super giants per spectral sub-type is to low for reliable statistics.}
              \label{lc600}
\end{figure*}

\begin{table*}
 \centering
\caption{Results of the fits for the one-to-one relation $Y(x)=Ax+a$ of the $A_{V}$ values for single stars using BVJHK magnitudes and intrinsic colours from Bessell et al. (1998, B98) and a combination from  Kenyon \& Hartmann (1995) and Schmidt-Kaler (1982) (KH95SK82).}
\label{phot_single}
\begin{tabular}{ccccc}

\hline
B98&		$A$&		95\% conf. intervall of $A$&		$a$&		95\% conf. intervall of $a$\\
\hline
B-V vs {\bf V-K}&	1.040&  		(1.016, 1.064)&	0.190&	  	(0.167, 0.213)\\	

B-V vs J-H&	1.175&  		(1.126, 1.225)&	-0.195&  	(-0.243, -0.147)\\

B-V vs J-K&	1.094&  		(1.048, 1.139)&	0.240&	  	(0.196, 0.284)\\	

{\bf V-K} vs J-H&	1.134&  		(1.099, 1.169)&	-0.414&  	(-0.456, -0.372)\\

\textbf{V-K} vs J-K&	1.097&  		(1.068, 1.126)&	0.002&	  	(-0.034, 0.037)\\	

J-H vs J-K&	0.847&  		(0.832, 0.862)& 0.470& 	 	(0.449, 0.490)\\	

\hline
KH95SK82&		$A$&		95\% conf. intervall of $A$&		$a$&		95\% conf. intervall of $a$\\
\hline

B-V vs V-K&	1.030&  		(1.006, 1.054)&	0.182&	  	(0.158, 0.205)\\

B-V vs V-J&	1.045& 		(1.021, 1.068)&	0.152&	  	(0.129, 0.175)\\

B-V vs V-K&	1.039&  		(1.015, 1.063)&	0.220&  		(0.197, 0.244)\\

\textbf{V-J vs V-H}&	0.998&  		(0.993, 1.003)&	-0.019&  	(-0.025, -0.012)\\

\textbf{V-J vs V-K}&	0.995&  		(0.988, 1.001)&	0.050&	 	(0.041, 0.058)\\

\textbf{V-H vs V-K}&	0.998&  		(0.994, 1.002)&	0.067&  		(0.062, 0.072)\\

\hline
KH95SK82 vs B98&		$A$&		95\% conf. intervall of $A$&		$a$&		95\% conf. intervall of $a$\\
\hline

B-V &		0.979&  		(0.972, 0.987)&	0.009&	  	(0.002, 0.017)\\

\textbf{V-K}&		0.977&  		(0.970, 0.985)&	-0.014&		(-0.024, -0.005)\\
\hline
\end{tabular}
\end{table*}

\noindent
Generally, the extinctions derived from the different authors and different magnitudes are well in agreement (see also Figures \ref{B_V}-\ref{KH}). The final $A_{V}$ value for each star is calculated from the mean of the $A_{V}$ values from the four colours marked bold in Table \ref{phot_single}. 109 stars have mean $A_{V}$ values below zero (probably due to variability and non-simultaneous photometry) but all of them are consistent to zero within their $1\sigma$ error. We set the $A_{V}$ values for these 109 stars to zero for all further calculations.\\
For the resolved 247 binary systems the various catalogues list spectral types and visual magnitudes for both components. We estimated the BJHK magnitudes of the components from the BVJHK magnitudes and the spectral types of the unresolved system listed in 2MASS and/or Simbad using the resolved V magnitudes and spectral types given in the catalogues of both components with a procedure as in Hohle et al. (2009).\\
From the resolved BVJHK magnitudes we calculate the $A_{V}$ values using the selected colours mentioned before.\\

\begin{table*}
\centering
\caption{List of the first ten from 2323 stars (see also Tab. \ref{initVal}). We derived the luminosities from the corrected (Smith \& Eichhorn, 1996) parallaxes. From luminosities and effective temperatures we calculated the masses (using the models below and taking the errors of the luminosities into account) with medians and standard deviation. The complete table will be available at the ADS data base in electronic form.}
\label{mass}
\begin{tabular}{cc|c|cccccc}
\hline
  & Hip & L & \multicolumn{5}{c}{mass}\\
&    &	[$L_{\odot}$]	&	\multicolumn{5}{c}{[$M_{\odot}$]}\\
&    &	              &	Bertelli et al. (1994) & Claret (2004) & Schaller et al. (1992) &	median & std. deviation\\
\hline

1 & 30122 & 3600 &	7.15-7.60 &	6.31-7.94 &	7.00 &	7.15 &	0.51\\
2 & 86414 & 2300 &	6.35-6.75 &	6.31 &	7.00 &	6.55 &	0.35\\
3 & 39138 & 1400 &	6.05-6.30 &	6.31 &	5.00-7.00 &	6.30 &	0.75\\
4 & 97278 & 2500 &	5.12-5.66 &	6.26 &	4.94-4.98 &	5.66 &	0.66\\
5 & 69996 & 2100 &	6.75-7.00 &	6.31 &	7.00 &	6.80 &	0.36\\
6 & 99473 &	848.6 & 4.18-4.63 & 3.98-5.01 & 4.00 & 4.00 & 0.35\\
7 & 76600 &	2704.9 & 7.25-7.65 & 7.94 & 7.00 & 7.25 & 0.49\\
8 & 67464 &	4428.8 & 8.50-8.80 & 7.94 & 9.00 & 8.50 & 0.53\\
9 & 79404 &	2504.2 & 7.30-7.90 & 7.94 & 7.00 & 7.60 & 0.48\\
10 & 32759 &	18900 & 10.95-12.90 & 9.97-12.52 & 8.97-11.94 & 11.94 & 0.49\\

\hline
\end{tabular}
\end{table*}

\begin{table}
\caption{From Table \ref{mass} we obtain typical error weighted median masses and luminosities for different spectral types and sub-types. We list these masses (together with the luminosities using corrected parallaxes from Figure~\ref{lc}) where at least five stars for one spectral sub-type are in the sample. Because of uncertain photometry we excluded binaries and all stars listed in Simbad with a range for their possible luminosity class. Note that the number of O stars in the sample is small and that they often have large distances (with large errors), i.e. their masses and luminosities are less reliable than for other spectral types.}
\label{mass_single}
\begin{tabular}{c|cc|cc|c}\hline
       & \multicolumn{2}{c}{mass} & \multicolumn{2}{c}{$L_{bol}$} & \# \\
SpType & median & std. dev.       & median & std. dev.                      &    \\
       & [$M_{\odot}$] &  [$M_{\odot}$] & [$L_{\odot}$] & [$L_{\odot}$]     &    \\

\hline

\multicolumn{6}{c}{LC I}\\

\hline

O9	&	24.25	&	5.80 & 146000 & 13000	&	 9\\
B0	&	15.00	&	2.62 & 27500 & 1000	&	27\\
B1	&	9.97	&	1.28 & 14300  & 300	&	55\\
B2	&	8.99	&	1.35 & 13400  & 100	&	40\\
B3	&	8.99	&	2.05 & 13800  & 1800	&	15\\
F5	&	7.53	&	2.69 & 6500 & 200	&	6\\
K3	&	6.26	&	1.64 & 3200  & 200	&	 9\\
M2	&	2.93	&	0.75 & 3900  & 800	&	 7\\

\hline

\multicolumn{6}{c}{LC III}\\

\hline

O9	&	17.77	&	7.00 & 43300 &	8600 &	 6\\
B0	&	13.75	&	3.37 & 18300 &	2300 &	16\\
B1	&	11.98	&	1.70 & 12200 &	600  &	42\\
B2	&	7.94	&	1.01 & 4900 &	 40  &	61\\
B3	&	6.31	&	0.72 & 2200  &	 40  &	68\\
B4	&	5.01	&	1.15 & 1300   &	 50  &	20\\
B5	&	5.00	&	0.51 & 800   &	 20  &	83\\
B6	&	4.65	&	0.72 & 580   &	 30  &	36\\
B7	&	4.00	&	0.59 & 490   &	 10  &	46\\
B8	&	4.00	&	1.23 & 530   &	 60  &	11\\

\hline

\multicolumn{6}{c}{LC V}\\

\hline

O7	&	17.52	&	9.33 &  13900 &  700 &	6\\
O9	&	19.60	&	4.33 & 36400 & 3100 &	13\\
B0	&	15.00	&	2.83 & 16100 &	130 &	27\\
B1	&	11.98	&	1.24 &  12520 &	150 &	81\\
B2	&	8.50	&	0.62 &  4130 &	 60 &	179\\
B3	&	6.55	&	0.42 &  1770 &	 20 &	219\\
B4	&	5.75	&	0.64 &  1260 &	 20 &	71\\

\hline
\end{tabular}
\end{table}

\section{Luminosities}

Since one does not measure the distance itself, but the parallaxe, we use the error dependent expectation values of the parallax, which lead to smaller distances. This treatment is introduced in Smith \& Eichhorn (1996). We apply this correction to all stars in our sample using equation 21 in Smith \& Eichhorn (1996). The errors of Hipparcos parallaxes are often as large as its value for a distance of $\geq1kpc$. 
Unfortunately OB-type stars are typically far from us. 1536 of the 2323 stars are within 600pc that is a reliable distance for Hipparcos. Due to the small relative errors for this distances the Smith \& Eichhorn (1996) correction does not strongly affect the distance estimate, while for stars with parallaxes $\approx1mas$ this effect becomes important, see Figure~\ref{dist} and~\ref{plxcorr}.\\
The luminosity of a star in units of $L_{\odot}$ can be calculated by this familiar equation: \\

\begin{figure}
  \centering
   \resizebox{\hsize}{!}
{
   \includegraphics[viewport=100 260 500 550, width=0.48\textwidth]{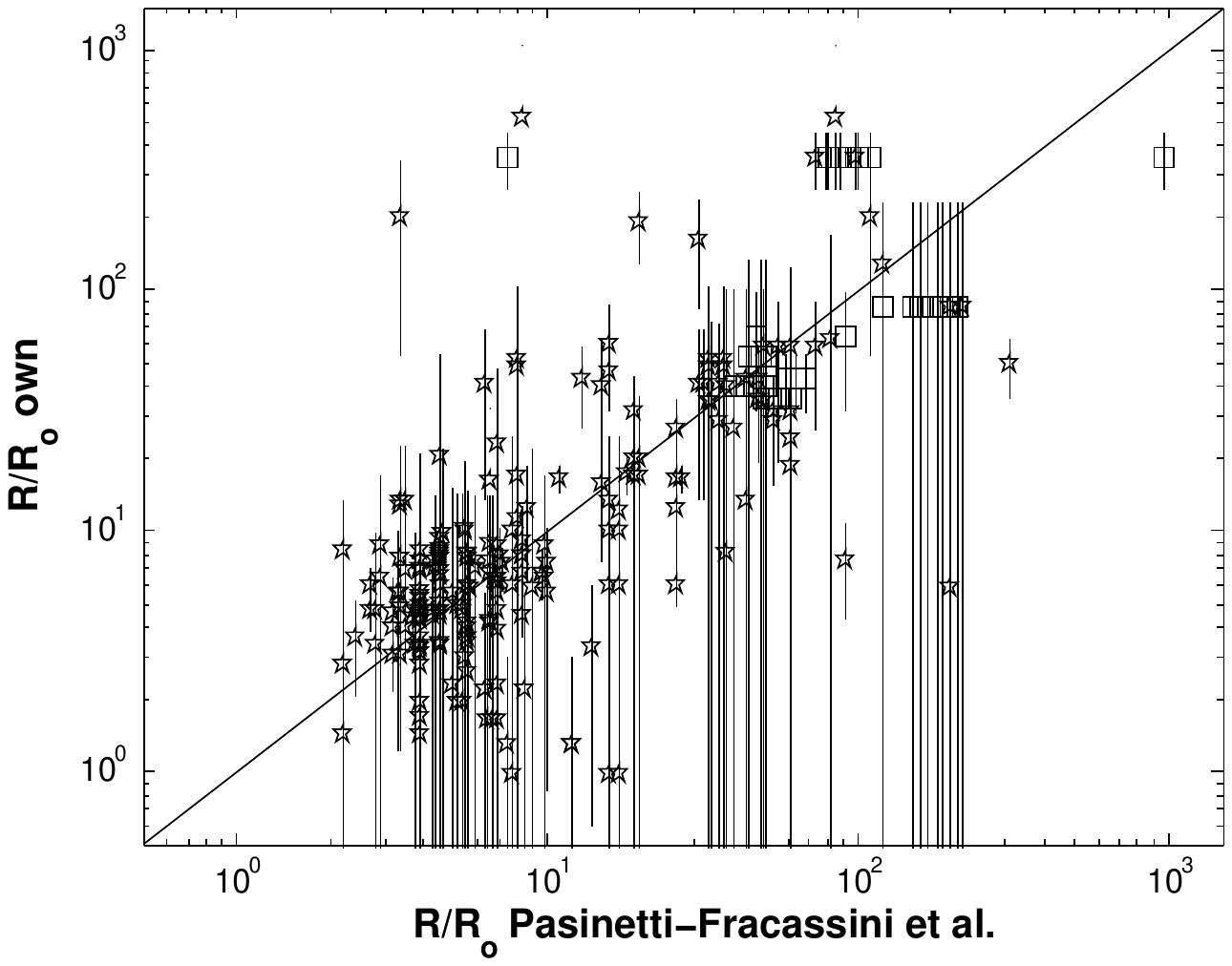}
}
   \caption{Own radii derived from our luminosities and temperature from the Stefan-Boltzmann law with $1\sigma$ error bars compared to radii from Pasinetti-Fracassini et~al. (2001) calculated from intrinsic brightness and colour (stars) and pulsating stars (squares). Our errors are mainly caused by the errors of the parallaxes.}
              \label{radSE}
    \end{figure}

 \begin{figure}
  \centering
   \resizebox{\hsize}{!}
{
   \includegraphics[viewport=100 260 500 550, width=0.48\textwidth]{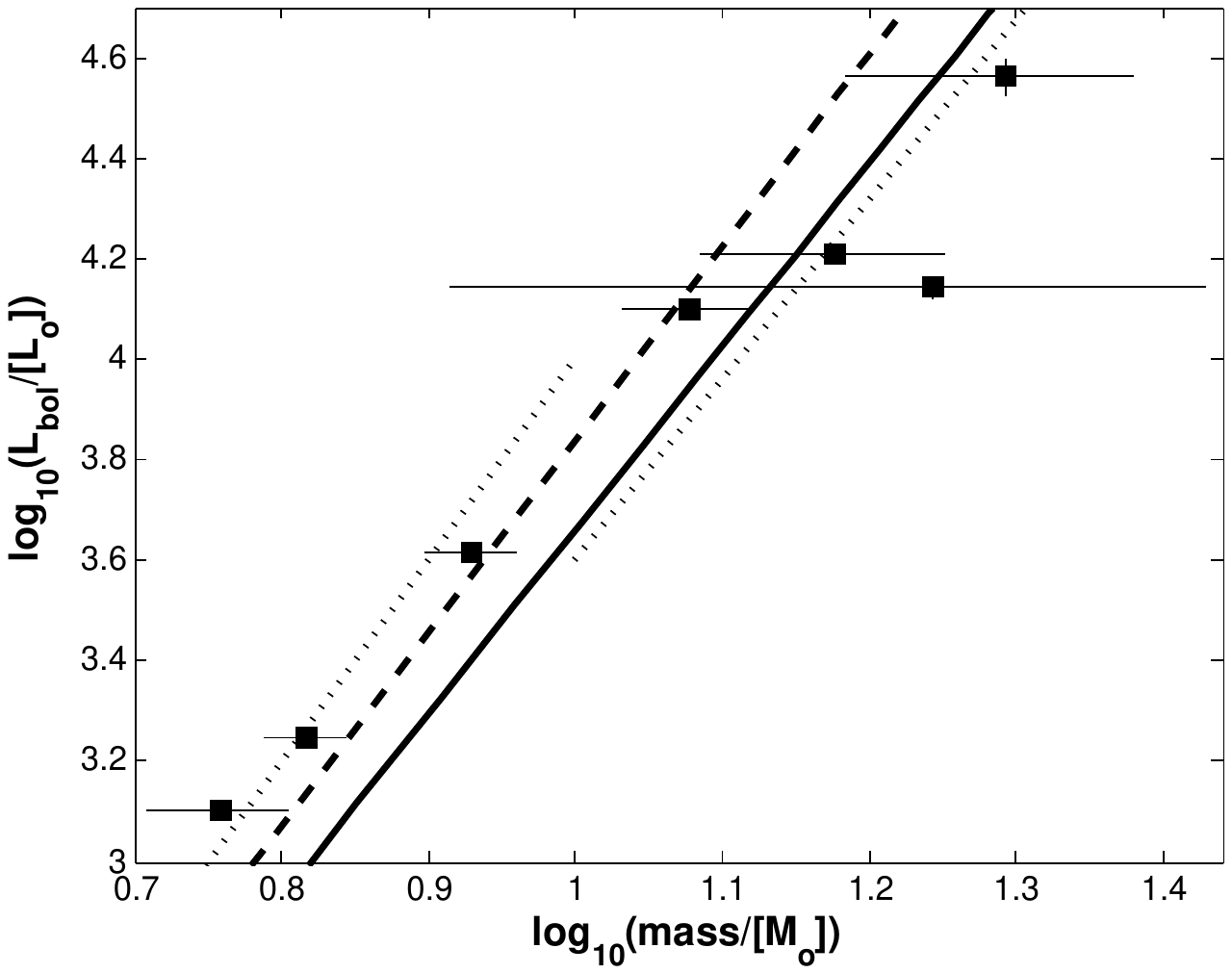}
}
   \caption{We derive a logarithmic slope of $3.66\pm0.12$ (black solid line) for the mass - luminosity relation for main sequence stars with data listed in Table~\ref{mass_single} (filled squares with $1\sigma$ error bars) that is slightly less than the slope of 3.84 (dashed line) obtained from the data in Andersen (1991, Table 1 therein). Hilditch (2001) gives a slope of 4.0 for stars with less than $10M_{\odot}$ and 3.6 for stars with larger masses (dotted lines).}
              \label{ML}
\end{figure}  

\begin{figure}[t]
  \centering
   \resizebox{\hsize}{!}
{
   \includegraphics[viewport=100 260 500 550, width=0.48\textwidth]{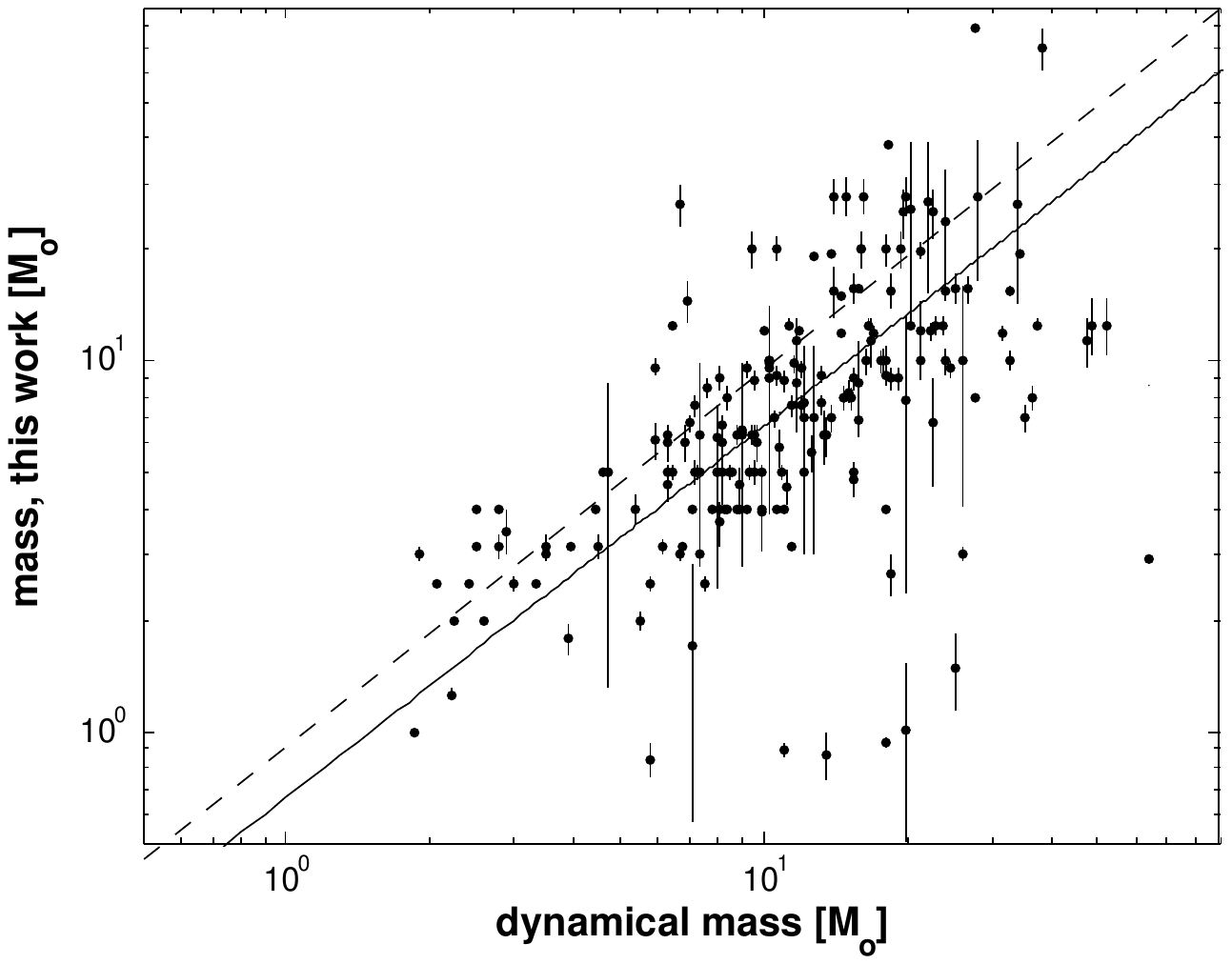}
}
   \caption{Masses of both components of binary systems using evolutionary models in this work derived from effective temperatures and luminosities (see Brancewicz \& Dworak, 1980; Bondarenko \& Perevozkina, 1996; Perevozkina, 1999; Surkova \& Svechnikov, 2004) compared to the dynamical mass values therein. Our mass values are medians from different models (see also Table \ref{mass}) with the standard deviations as errors. The dashed line indicates the 1:1 relation: Our masses underestimate the dynamical masses of a factor of 1.5 in median (solid line).}
              \label{massaut}
    \end{figure}   

\begin{equation}
  L_{bol}=10^{0.4(5log_{10}d-5+4.74-BC_{V}-m_{V}+A_{V})} 
\end{equation}

\noindent with the corrected distance d in parsec.\\ 
From spectral types we derived the temperatures $T_{eff}$ using the Tables in Bessell et al. (1998), Kenyon \& Hartmann (1995) and Schmidt-Kaler (1982). With temperature and luminosity we show all 2323 stars, singles and massive binaries, in the H$-$R diagram in Figure \ref{HRD}.
Our sample contains dozens or even more than hundred stars for most spectral types. This enables us to provide reliable statistic median luminosities with standard deviations for each spectral type (if the given sub-type is not an integer number, for example B2.5, we round up to the later spectral type given the slope of the temperature to spectral type conversion). We compare in Figure \ref{lc} our median luminosities with at least five stars per spectral sub-type with previously published (standard) bolometric luminosities from Schmidt-Kaler (1982) listed in Lang (1994).\\

While for most spectral types the values from Schmidt-Kaler (1982) (who do not list errors) are consistent with ours, there is a tendency to smaller luminosities for LC I and III in our new data. This discrepancy is still present if we restrict our statistics to stars within 600pc (corrected distances) or use uncorrected parallaxes for the luminosities (because most of the stars are within 600pc where the Smith \& Eichhorn (1996) correction is not important, see Figures~\ref{lc} and~\ref{lc600}).\\
Wegner (2007) calculated luminosities from a star sample which is quite similar to ours, but shows only spectral types later than A0, with Hipparcos parallax (but extinctions only from B-V, not from BVJHK) and compared the result to the luminosities from Schmidt-Kaler (1982). Wegner (2007) found for late type super giants the same differences as we: they are under luminous compared to Schmidt-Kaler (1982) by 1.5 magnitudes in average, in particular also the large discrepancy around spectral type K and M (up to two magnitudes).\\ 
Recently, photometric distances of 29 OB associations and many OB field stars were adjusted using Hipparcos parallaxes from Dambis et al. (2001). They found, that previous photometric distances overestimated the Hipparcos distances about 11\% on average for these OB associations and about 20\% for field stars, respectively.\\
With corrected luminosity and effective temperature we can estimate the stellar radius with the Stefan-Boltzman law.
We show the stellar radii from those stars of our sample which are listed in Pasinetti-Fracassini et al. (2001) and our own radii in Figure \ref{radSE}. Pasinetti-Fracassini et al. (2001) provide a list of radii measured directly with different methods between 1950 and 1997 (681 values for 246 stars). We see a good consistency of our radii to those of Pasinetti-Fracassini et al. (2001).

\section{Masses}

\begin{table*}
\caption{List of 36 binary systems where both components exceed 8$M_{\odot}$ from dynamical masses given in the binary catalogues discussed in the text (see also column seven). We list mass and spectral type ranges obtained from these catalogues for both components.}
\label{mass_bin}
\begin{tabular}{cc|cc|cc|r}

\hline

 	&	Hip	& \multicolumn{2}{c|}	{SpType} &	\multicolumn{2}{c|}{mass/[$M_{\odot}$]} &	ref. \\
  &     & primary & secondary & primary & secondary & \\
\hline

1	& 1415  &	O7/O9III & O8-9/O9III	& 20.30-57.75 & 14.8-31.73 & [1], [2], [4]\\						
2	& 4279  &	A5I & G0I & 19.88 & 9.94 & [2] \\
3	&	15063 &	O9.3IV & O9IV & 20.37 & 9.98 & [2], [4] \\
4	&	25565 &	B0V/O9.5V & B1/B2/B0.5 & 12.04-21.30 & 7.95-14.50 & [1], [2], [4] \\
5	&	25733 &	O9.5/O9.5III & B0IV/09.5III & 21.28-24.00	& 12.7-18.90 & [1], [2], [4] \\
6	&	28045 &	B4V/F3eIb	& K5II & 18.46 & 11.08 & [2], [4] \\         		
7	&	29276 &	B3III/B1V/B0.5III & B3.5/B3V & 15.48-16.90 & 8.51-9.00 & [1], [2], [4] \\
8	&	31939 &	B1.5IV & B3 & 11.50 & 8.40 & [1] \\        		
9	&	33953 &	B2.5IV/B2.5IV-V & B2.5IV-V & 15.35 & 15.35 & [2], [4] \\
10	&	34646 &	B3 & B4 & 11.10	& 8.88 & [2] \\        		
11	&	35412 &	O7 & O9III/O7.5& 22.00 & 18.30 & [1], [2], [4] \\
12	&	56196	&	B5-O7 & B8-O9.5	 & 8.24-22.60 & 7.75-15.40 & [2], [4], [5]	\\
13	&	57895	&	B1III & ? & 14.58 &	10.21	& [2]	\\
14	&	59483	&	G2I	& ? & 11.67 & 8.17 &  [2]	\\
15	&	85985	& B1V & B1.5 & 10.30 & 10.20	& [1]		\\
16	&	89769	&	WC7-8	& B0/O8-9III-V & 18.49 & 	11.28 & [2], [4]  \\
17	&	92055	&	B3 & B3 &	22.39 & 15.01 & [2]			\\
18	&	92865 &	O9/O9V & B1-3/B3V & 18.01-38.20 & 10.81-13.80 & [1], [2], [4] \\
19	&	93502 &	B2/B3.5/B4V & B3.5-8 & 18.03-18.40 & 11.36-11.40 & [1], [2], [4] \\         		
20	&	95176	&	A5I & M5Ia & 25.18 & 19.14 & [2]	\\
21	&	97634 &	B1.5II-III/B1III & B2-3V & 16.70 & 9.35	& [2], [4]  \\        
22	&	99021 &	O9.5e/O9.5V/O8.9V& B1I-II/B1Ib/B1.2Ib	& 23.84-25.20 & 14.00-15.73 & [2], [3]  \\
23	&	100135 & O6.5/O6.5V/O7.5 & O7.5/O9 & 28.00-37.16 & 19.60-32.70 & [1], [2], [4]  \\ 
24	&	100193 & B2 & B2/B2.5 & 13.82 & 12.16	 & [2], [4]  \\ 
25	&	100214 & WN5-5.5 & B1/O8III &	34.53 & 19.34	& [2], [4] 	\\
26	&	101341 & O7/O7f  & O9-B0/O8	& 26.70-27.80  & 6.70-22.96	 & [1], [2]  \\
27	&	102648 & A5Iab/A5epIa & A9	& 12.62 & 8.83 & [2], [4]  \\        		
28	&	102999 & B0IV	& B0IV 	& 17.71	& 17.53	 & [2], [4]  \\  
29	&	103419 & K5I & B4V & 	22.64	 & 	8.1504 &  [2]		\\
30	&	108073 & B0.5V & B1V	& 10.51	& 9.46 & [2]  \\         		
31	&	108317 & M2epIa & B8Ve/B9  & 63.81	 & 35.10 & [2], [4]  \\
32	&	110154 & WN6 & B0III & 23.95 &	16.05 & [2]  \\
33	&	112470 & O5 & O5 & 34.00 & 27.70 & [1]  \\
34	&	112562 & B0.5V/O8 & B0.5V/B0.5/O9 & 15.22-18.10 & 13.24-15.90 & [1], [2], [4] \\       		
35	&	113461 &	B0IV & B0IV	& 16.07 & 13.98 &  [2],	[4]	 \\
36	&	113907 & B0.5/B0.5IV-V & B0.5IV-V & 10.62 & 9.45 & [2], [4]  \\ 
\hline

\multicolumn{7}{l}{[1] Bondarenko \& Perevozkina (1996)}\\
\multicolumn{7}{l}{[2] Brancewicz \& Dworak (1980)}\\
\multicolumn{7}{l}{[3] Surkova \& Svechnikov (2004)}\\
\multicolumn{7}{l}{[4] Pourbaix et al. (2007)}\\    
\multicolumn{7}{l}{[5] Docobo \& Andrade (2006)}\\      
     
\hline
\end{tabular}
\end{table*}

With luminosities and effective temperatures, we can estimate the masses of our stars by
comparing their location in the H-R diagram with model mass tracks. We use evolutionary models from Schaller et al. (1992), Bertelli et al. (1994) and Claret (2004). The different authors provide evolutionary tracks for different metallicities, we present the masses for solar metallicity. The metallicity is only well known for few stars and affect the mass estimation only by a few percent. The differences in mass between the models with the same metallicities are comparable to this.\\
Owing to the discrepancies in the luminosities for super giants, we fix the temperature first, that is much better known than the luminosity, within 10\% tolerance taking possible uncertainties from the spectral type determination into account. We then determined the mass tracks with the best relative agreement to the given luminosities within their errors. Even if the luminosity is strongly underestimated and the relative deviation to the nearest mass track will be large, at least it will be on the main sequence. This avoids a systematic underestimation of the masses caused from underestimated luminosities.\\
The models of Schaller et al. (1992) underestimate the masses by 0.37\% in median compared to the masses obtained from Bertelli et al. (1994), while Clarets (2004) model overestimates the masses about 2.7\% in median compared to the masses from the model of Bertelli et al. (1994), i.e. they agree well. While Claret (2004) and Schaller et al. (1992) provide models for masses up to $120M_{\odot}$, the models from Bertelli et al. (1994) give masses up to $34M_{\odot}$ for solar metallicity. Therefore, if the mass estimation of a star using Schaller et al. (1992) or Claret (2004) results in $34M_{\odot}$ or more, we did not use the results using Bertelli et al. (1994). We list the results of the first ten stars in Table \ref{mass}.\\
Using the different results from the different models for each star, we find, that the mean of the standard derivation is 9.9\% compared to the median of the masses themselves. We see this as good consistency. For 76\% of the stars, the standard deviation of the mass is less than 10\% of the median of the mass value and for 28\% of the stars the standard deviation is less than 5\%. The standard deviations of the mass values may underestimate the error of mass estimation.\\
Having determined the masses of all 2323 stars, we obtain median masses for the spectral sub-types depending on the LC. Likewise for the bolometric luminosities we have dozens, or even more than hundred, stars per spectral sub-type, i.e. the median masses should be robust against fluctuations and errors in the empirical data. We list these masses for stars with at least five entries in a spectral sub-type in Table \ref{mass_single}. If a system appears in one of the used binary catalogues, we use its dynamical mass instead of our model-dependent masses. If one system appears in more than one of these catalogues, we use the median and the standard deviation (as error) from the different mass values.\\
From masses and luminosities in Table \ref{mass_single} we derive a mass - luminosity relation ($L \propto M^{\beta}$) with $\beta=3.66\pm0.12$ for the main sequence stars (see Figure~\ref{ML}).\\
We also compare the masses of the binaries with dynamical masses with our method of mass determination. We use the effective temperatures and luminosities (derived from $M_{bol}$) listed in Brancewicz \& Dworak (1980), effective temperatures from listed spectral types and luminosities (derived from $M_{bol}$) in Bondarenko \& Perevozkina (1996), Perevozkina (1999) and Surkova \& Svechnikov (2004) to calculate own mass values (if a system appears in more than one catalogue we list the median of the different masses). Our masses are in good agreement but tend to smaller values (a factor of 1.5 in median, peak at $\approx 10-20\%$) compared to the masses from the other authors (see Figure \ref{massaut}).\\

\begin{figure*}
  \centering
   \resizebox{\hsize}{!}
{
   \includegraphics[viewport=25 100 700 550, width=0.48\textwidth]{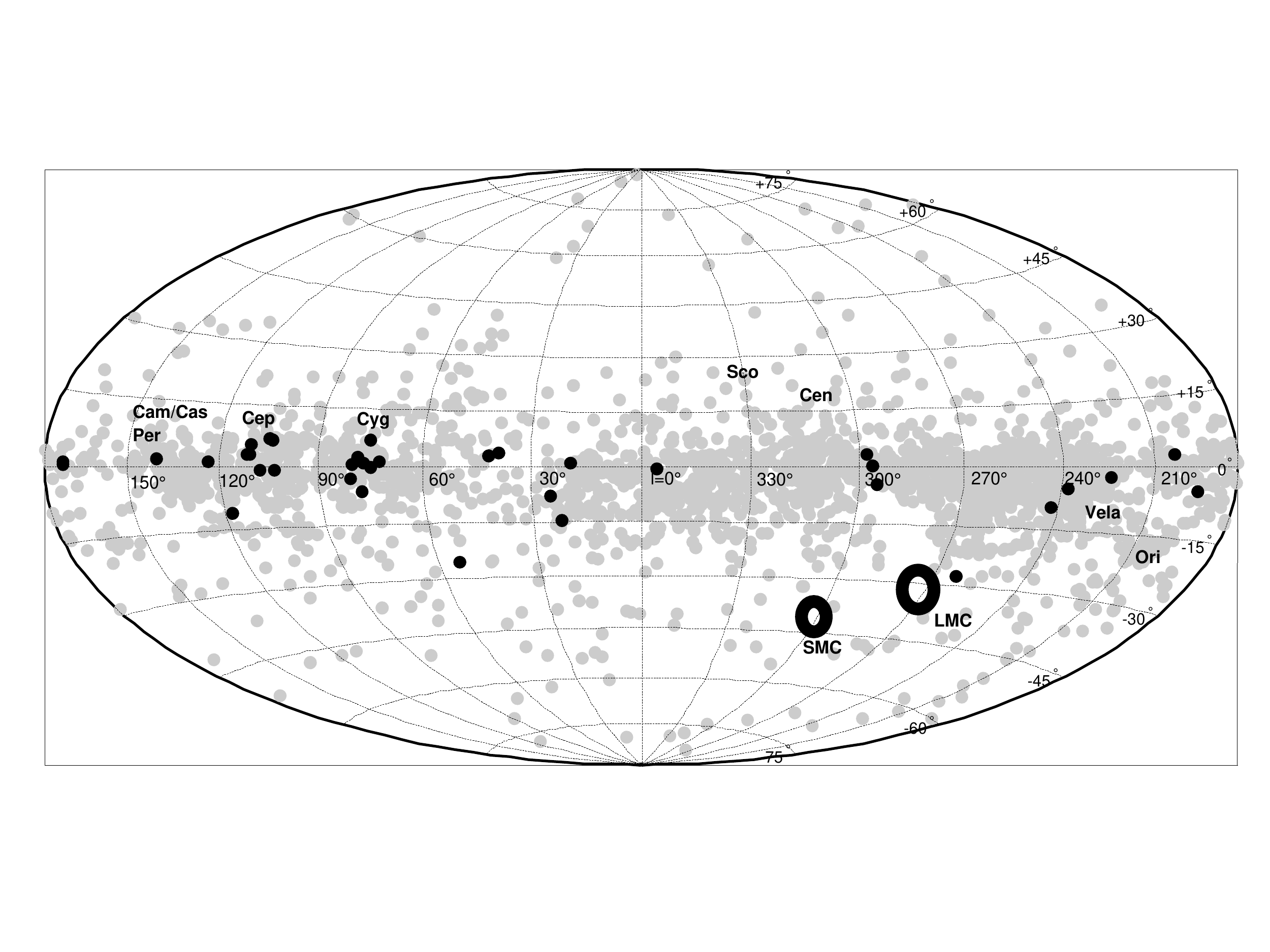}
}
   \caption{The complete star sample used in this work is represented by grey dots. Massive binaries are shown as black dots (with both components having masses $\geq8M_{\odot}$) from dynamical masses listed in Brancewicz \& Dworak (1980), Bondarenko \& Perevozkina (1996), Perevozkina (1999) and Surkova \& Svechnikov (2004). We indicate a few OB associations, where we find clusters of massive stars, i.e. predict more supernovae in the near future (see Figure~\ref{galSN_rate}). Both Magellanic Clouds (LMC and SMC) are indicated, which were left out for this work.}
              \label{NSbin}
    \end{figure*} 

\section{Super nova progenitors}

We find 759 stars in our sample with median masses $\geq8M_{\odot}$ in total, 36 of them are the secondaries of a more massive primary (Table~\ref{mass_bin}). Among them, in three systems the primary has a median mass $\geq30M_{\odot}$, i.e. may form a black hole. We list current masses in Table~\ref{mass_bin}, but do not include binary interaction for predicting the final outcome.\\ 

\begin{figure}[t]
  \centering
   \resizebox{\hsize}{!}
{
   \includegraphics[viewport=100 260 500 550, width=0.48\textwidth]{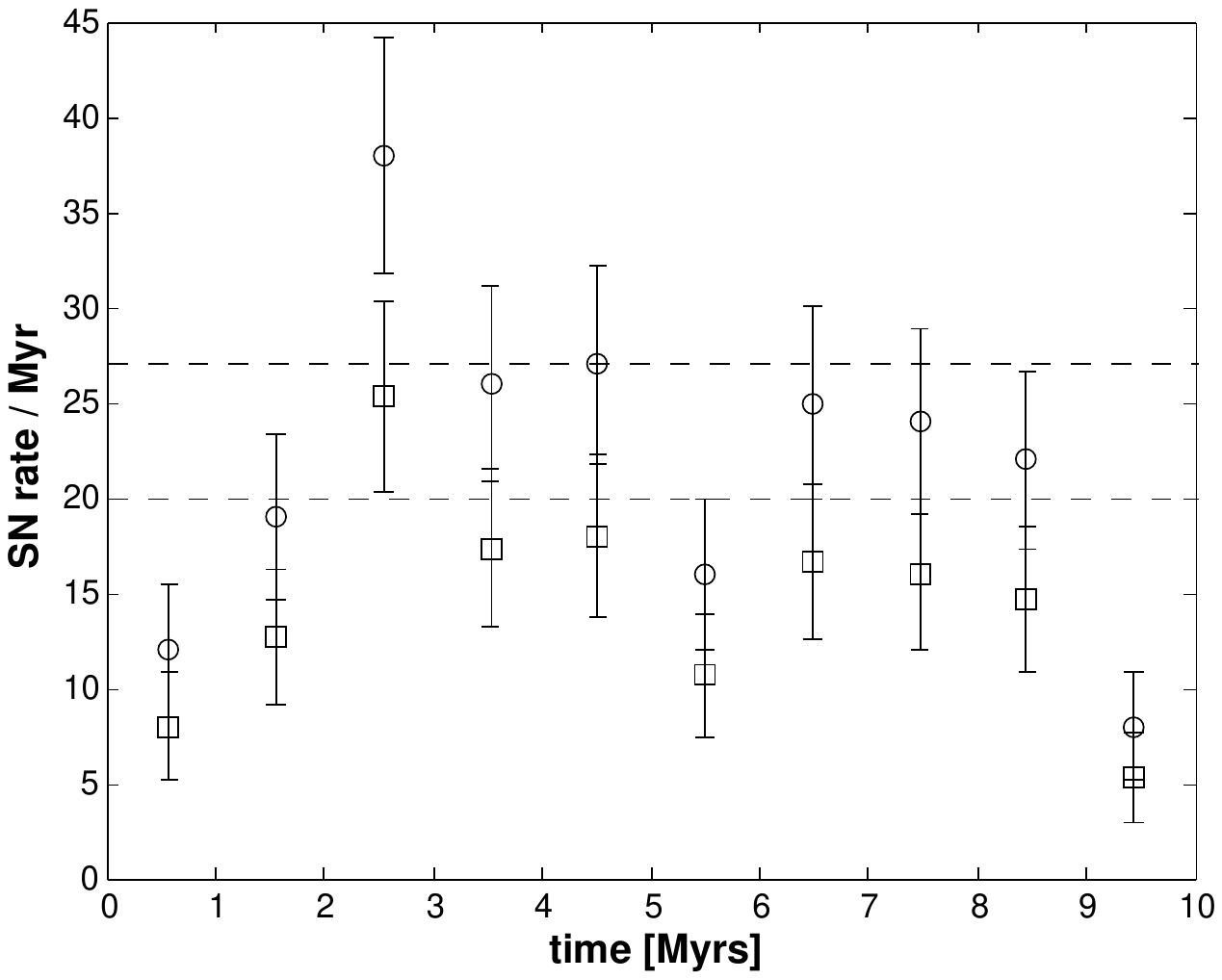}
}
   \caption{The supernova rate within 600~pc in the future (circles) obtained from the stars of our sample and multiplied with 2/3 (squares) for the Gould Belt rate (see text) compared with the rate of 20-27 events/Myr given in Grenier (2000) for the Gould Belt (dashed lines). Due to our selection criteria (see text) for the star sample, our rate is a lower limit. The average rate over 10Myrs is $\geq$14.2$\pm$3.8~events/Myr for the Gould Belt and $\geq$21.3$\pm$4.6~events/Myr within 600~pc. All errors are Poissonian.}
              \label{SN_rate}
\end{figure}

\begin{figure*}
  \centering
   \resizebox{\hsize}{!}
{
   \includegraphics[viewport=25 20 700 550, width=0.48\textwidth]{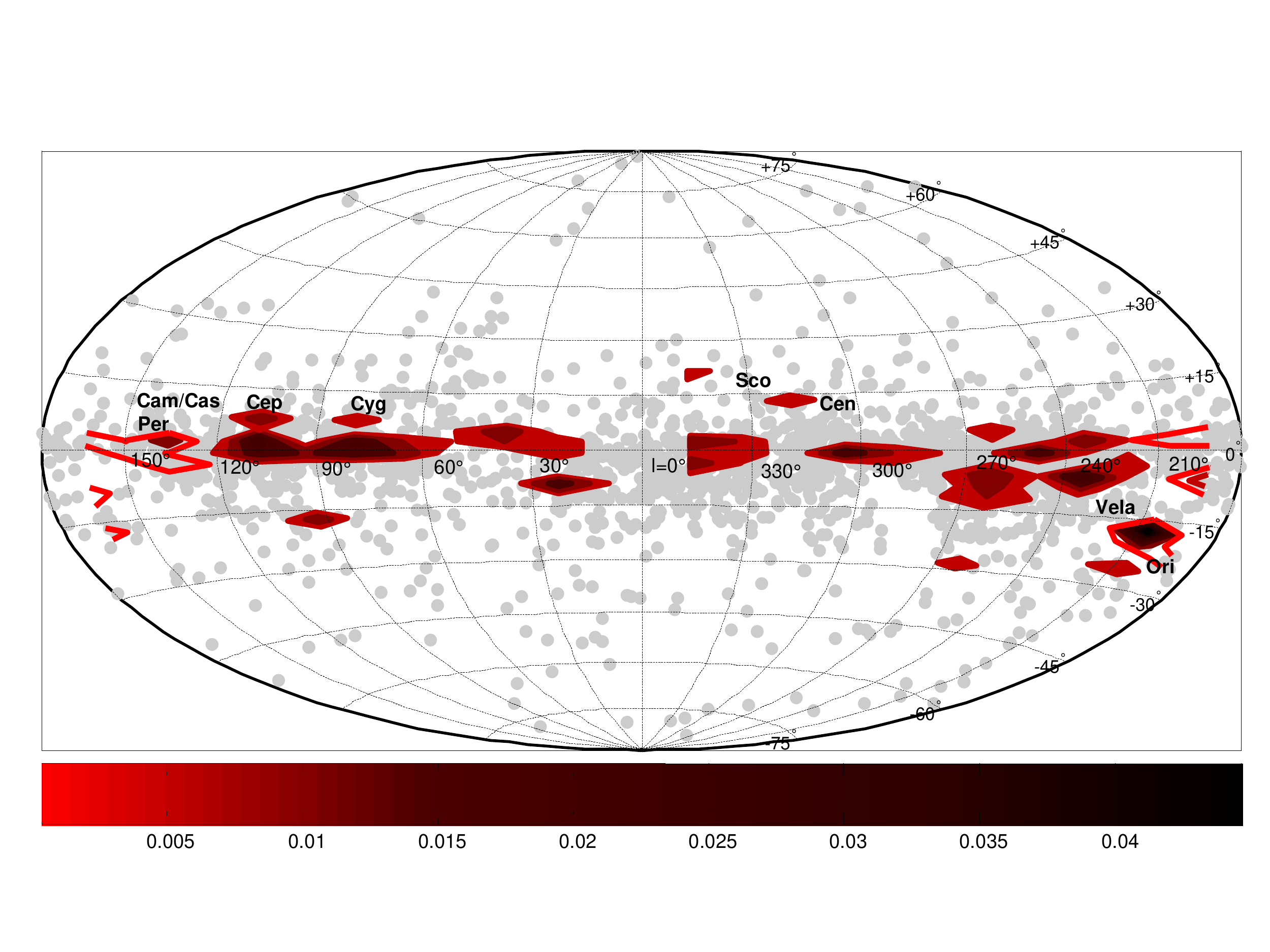}
}
   \caption{Same as in Figure~\ref{NSbin}, shown with the distribution of the super nova rate for the next 10~Myrs (see also Figure~\ref{SN_rate}) including all massive stars within 600pc. The colours indicate the normalised rate per Myr and square bin (longitude and latitude both divided in 25 bins). Note, that the super nova rate per area in Orion is 2-3 times higher than for the other clusters.}
\label{galSN_rate}
\end{figure*}

Starting from the median mass values with the standard deviations ($1\sigma$) as errors we can give a maximum number of such systems (median + $1\sigma$), a median number (median masses) and a minimum number (median - $1\sigma$), see Table \ref{sys}. We also give the corresponding numbers for these progenitors within 600pc in Table \ref{sys}. This includes the Gould Belt that hosts 2/3 of the SN progenitors within this distance (Torra et~al., 2000).\\   
From the mass estimation we also obtain ages using the corresponding isochrones in the models. Given masses and ages we estimate the expected remaining life time of a star using the model in Maeder \& Meynet (1989) and, hence, predict a SN rate for the near future (that should be similar to the SN rate of the recent past). This SN rate is stable until $\sim$10Myrs in the future, then star formation and evolution matter. We obtain a SN rate of 21.3$\pm$4.7~events/Myr in average (given the Poissonian error) within 600pc, i.e. 14.5$\pm$3.8~events/Myr for the Gould Belt. We show the SN rate for the next 10Myrs within 600pc and for the Gould Belt in Figure~\ref{SN_rate}.\\
We stress that we restrict our sample to those massive stars within 3kpc, which have both Hipparcos parallaxes and 2MASS JHK data, in order to estimate precise and accurate luminosities and masses. Hence, we miss several SN progenitor stars. Starting from 3694 stars (see Section 2) to 2323 stars in the final sample, we systematicly underestimate the number of such systems at least by a factor of 1.2 within 600pc and more than 1.6 for stars within 3kpc. Therefore we only estimate the SN rate for the well investigated and more complete stars within 600pc and multiply it with a factor of 1.2 (that still gives a lower limit of the rate), obtaining 17.4$\pm$4.2~events/Myr for the Gould Belt close to the past SN rate of the Gould Belt in Grenier (2000), also averaged over 10Myrs. The SN events are concentrated in OB clusters, in particular the Orion OB clusters and the Vela region (see Figure~\ref{galSN_rate}).

\begin{table}
\centering
\caption{Number of systems with at least one neutron star (NS) or black hole (BH) progenitor in the total sample (see text) and within 600pc in parenthesis plus the corresponding number if both components of a binary system are NS and/or BH progenitors. The numbers are obtained from the median mass values (median numbers), the minimum (median mass value - $1\sigma$) and the maximum (median mass value + $1\sigma$) numbers of progenitors.}
\label{sys}
\begin{tabular}{cccc}
\hline
& minimum & median & maximum\\
\hline
NS prog. within 3kpc & 686 + 26 & 759 + 36 & 1004 + 43 \\
				 within 0.6kpc & (287 + 12) & (356 + 19) &	(485 + 25) \\
\hline
BH prog. within 3kpc & 12 + 1 & 24 + 1 & 54 + 3 \\
				 within 0.6kpc & (2 + 0) & (2 + 0) & (8 + 0) \\
				 
\hline
\end{tabular}
\end{table}

\section{Conclusions}

Our mean luminosities and masses are derived from dozens of stars for most spectral types, which should make our results reliable and robust against individual outliers. In our sample 1536 stars are within 600pc and 2127 stars are within 1kpc. For most spectral types and luminosity classes, our luminosities are smaller than those in Schmidt-Kaler (1982), in particular for super giants, even if we restrict our luminosities to those stars with values of the parallax larger than its 3$\sigma$ error or to stars closer than 600pc.\\
This has several reasons:\\ 
\\
\begin{enumerate}
\item Hipparcos distances are smaller than previously used ground based distances. This effect of 20 - 30\% in distance results in a revision of luminosity of 44 - 70\%. We thereby confirm previous similar conclusions by Dambis et al. (2001) and Wegner (2007).
\item Many stars, especially super giants, which were supposed to be single stars decades ago, are now known as multiple or double systems with their components on the main sequence. Schmidt-Kaler (1982) uses data from Code et al. (1976). 75\% of the stars in Code et al. (1976) are known to be binaries or multiples today, but listed as single stars in Code et al. (1976).\footnote{One of those stars is HD 68273, which was known as WC8 + O9I (in Code at al., 1976) and is now known as O9 + B3 + A0 + A0 (CCDM catalogue, {\bf C}atalogue of the {\bf C}omponents of the {\bf D}ouble and {\bf M}ultiple stars, Dommanget \& Nys, 2000). The magnitudes $M_{V}(WC8)$ and $M_{V}(O9I)$ were measured as $(-4.8\pm0.3)mag$ and $(-6.2\pm0.2)mag$, respectively, from Conti \& Smith (1972). The distance was assumed to be 460pc in Abt et al. (1976) but was revised to 258pc in van der Hucht et al. (1997) using the Hipparcos parallax. The new distance yields to $M_{V}(WC8)=-3.7mag$ and $M_{V}(O9I)=-5mag$.} 
\end{enumerate}            
 
The masses we derived from our new luminosities using evolutionary models agree well with dynamical masses. We find 36 binaries with both components $\geq8M_{\odot}$ and estimated the SN rate for the next 10Myrs for the solar neighbourhood to be about one SN per 50kyr. We have restricted our sample here to those massive stars within 3kpc, which have both Hipparcos parallax and 2MASS JHK data. We will enlarge our sample including all possible super nova progenitors within 3kpc in further work.\\
Information about the likely distribution of neutron stars in the solar neighborhood can be important for the design of searches for gravitational waves (GWs) with current interferometric detectors like GEO600, LIGO and VIRGO. Blind searches for previously unknown neutron stars radiating gravitational waves are computationally very expensive, so restriction of searches to specific regions of the sky, frequencies, and spin-down time-scales can improve the sensitivity of searches. Of particular interest in current searches are old, isolated neutron stars which have cooled down so that they are no longer visible as X-ray sources, and which might not be radio pulsars or might have pulsar beams that are not directed toward us. Taking high kick velocities into account, 140 neutron stars, younger than 4 Myrs, should be still present within 1~kpc (Palomba, 2005).\\
Current GW searches for isolated neutron stars contain a spin-down parameter, which means that they can also detect accelerating systems, such as sources in wide binary systems. GW searches could easily be generalized to find neutron stars in wide binaries, even potentially those with accretion that leads to increased ellipticity and spin-up rather than spin-down.\\

\acknowledgements
The work is supported by the Deutsche Forschungsgemeinschaft (DFG) through
SFB/TR 7 "Gravitationswellenastronomie".\\
We would like to thank B. Allen, B. Owen, K. Schreyer, B. Posselt and A. Seifahrt for helpful discussions.
\newpage


\begin{thebibliography}{}
 
  \bibitem[1976]{Abt} Abt, H. A., Landolt, A. U., Levy, S. G.; Mochnacki, S., AJ, 81, 541  
 \bibitem[1991]{Andersen} Andersen, J., 1991, A\&ARv, 3, 91  
  \bibitem[1994]{Bertelli} Bertelli, G., Bressan, A., Chiosi, C., Fagotto, F., Nasi, E., 1994, A\&A, 106, 275
  \bibitem[1998]{Bessell} Bessell, M. S., Castelli, F., Plez, B., 1998, A\&A, 333, 231 
  \bibitem[1996]{Bondarenko} Bondarenko, I.~I. Perevozkina, E.~L., 1996, Odessa Astronomical Publications, 9, 20
  \bibitem[1980]{Brancewicz} Brancewicz, H.~K., Dworak, T.~Z., 1980, Information Bulletin on Variable Stars, 1744, 1  
   \bibitem[2004]{Claret} Claret, A., 2004, ApJ, 417, 434
  \bibitem[1976]{Code} Code, A. D., Davis, J., Bless, R. C., Brown, R. H., 1976, A\&A, 424, 919     
  \bibitem[1972]{Conti} Conti, P. S., Smith, L. F., 1972, ApJ, 172, 623 
  \bibitem[2003]{Cutri} Cutri, R. M., Skrutskie, M. F., van Dyk, S., et al., 2003, The IRSA 2MASS All-Sky Point Source Catalog, NASA/IPAC Infrared Science Archive 
  \bibitem[2001]{Dambis} Dambis, A. K., Mel'nik, A. M., Rastorguev, A. S., 2001, Astronomy Letters, 27, 58
  \bibitem[2006]{Docobo} Docobo, J.~A., Andrade, M., 2006, ApJ, 652, 681
  \bibitem[2000]{Dommanget} Dommanget, J., Nys, O., 2000, O\&T, 52, 26
  \bibitem[2000]{Grenier} Grenier, I. A., A\&A, 364, 93
  \bibitem[2001]{Hilditch} Hilditch, R. W., 2001, icbs.book
  \bibitem[2009]{Hohle} Hohle, M. M.; Eisenbeiss, T.; Mugrauer, M. et al., 2009, AN, 330, 511
  \bibitem[1995]{Kenyon} Kenyon, S. J., Hartmann, L., 1995, ApJS, 101, 117 
  \bibitem[1994]{Lang} Lang, K. R., 1994, A\&A, 105, 39
  \bibitem[1989]{Maeder} Maeder, A., Meynet, G., 1989, A\&A, 210, 155
  \bibitem[2005]{Palomba} Palomba, C., 2005, MNRAS, 359, 1050
  \bibitem[2001]{Pasinetti} Pasinetti-Fracassini, L. E., Pastori, L., Covino, S., Pozzi, A., 2001, A\&A, 367, 521
  \bibitem[1999]{Perevozkina} Perevozkina, E.~L., 1999, VizieR Online Data Catalog 
  \bibitem[2003]{Perrot} Perrot, C. A., Grenier, I. A., 2003, A\&A, 404, 519
  \bibitem[1997]{Perryman} Perryman, M. A. C., Lindegren, L., Kovalevsky, J. et al., 1997, A\&A, 323, 49 
  \bibitem[2005]{Popov} Popov, S.B., Turolla, R., Prokhorov, M. E., Colpi, M., Treves, A., 2005, Ap\&SS, 299, 117
  \bibitem[2007]{Pourbaix} Pourbaix, D., Tokovinin, A.~A., Batten, A.~H., et al., 2007, VizieR Online Data Catalog  
  \bibitem[1982]{Schmidt} Schmidt-Kaler, T. H., 1982, Landolt-Bornstein New Series, Volume 2b, Springer Verlag, New York (82)
  \bibitem[1992]{Schaller} Schaller, G., Schaerer, D., Meynet, G., Maeder, A., 1992, A\&AS 96, 269
  \bibitem[1996]{Smith} Smith, H., Eichhorn, H., 1996, MNRAS, 281, 211 
  \bibitem[2004]{Surkova} Surkova, L.~P., Svechnikov, M. A., 2004, VizieR Online Data Catalog   
  \bibitem[2000]{Torra} Torra, J., Fernandez, D., Figueras, F., 2000, A\&A, 359, 82
 \bibitem[1997]{Hucht} van der Hucht, K. A., Schrijver, H., Stenholm, B., Lundstr\"om, I., Moffat,  A. F. J., Marchenko, S. V. , Seggewiss, W., Setia Gunawan, D. Y. A., Sutantyo W., van den Heuvel,  E. P. J., De Cuyper,  J. P., Gómez,  A. E., 1997, NewA, 2, 245V 
  \bibitem[2007]{Leeuwen1} van Leeuwen, F., 2007a, A\&A, 474, 653 
  \bibitem[2007]{Leeuwen2} van Leeuwen, F., 2007b, Hipparcos, the new reduction of the raw data, Dortrecht:Springer 
  \bibitem[2007]{Wegner} Wegner, W., 2007, MNRAS, 374, 1549
  \bibitem[2007]{Zinnecker} Zinnecker, H., Yorke, H. W. 2007, ARA\&A, 45, 481
\end{thebibliography}
\end{document}